%% file: conference_101719.tex
\documentclass[conference]{IEEEtran}
\IEEEoverridecommandlockouts
\usepackage{cite}
\usepackage{amsmath,amssymb,amsfonts}
 \usepackage{algorithm}
\usepackage[noend]{algpseudocode}
\usepackage{graphicx}
\usepackage{textcomp}
\usepackage{xcolor}

\def\BibTeX{{\rm B\kern-.05em{\sc i\kern-.025em b}\kern-.08em
    T\kern-.1667em\lower.7ex\hbox{E}\kern-.125emX}}
\begin{document}

\title{DPCOVID: Privacy-Preserving Federated Covid-19 Detection\\
}

\author{\IEEEauthorblockN{1\textsuperscript{st} Trang-Thi Ho}
\IEEEauthorblockA{\textit{Research Center for Information Technology
Innovation} \\
Academia Sinica, Taipei, Taiwan \\
hothitrang.dhdn@gmail.com}
\and
\IEEEauthorblockN{2\textsuperscript{nd} Yennun-Huang}
\IEEEauthorblockA{\textit{Research Center for Information Technology
Innovation} \\
Academia Sinica, Taipei, Taiwan\\
yennunhuang@gmail.com}
}

\maketitle

\begin{abstract}
\input{abstract}
\end{abstract}


\section{Introduction}
\label{introduction}
\input{introduction}

\section{Background}
\label{background}
\input{background}

\section{Experiments}
\label{experiments}
\input{experiments}

\section{Conclusion}
\label{conclusion}
\input{conclusion}

\bibliographystyle{IEEEtran}
\bibliography{IEEEabrv,mybibfile}

\end{document}

%% file: abstract.tex

Coronavirus (COVID-19) has shown an unprecedented global crisis by the detrimental effect on the global economy and health. The number of COVID-19 cases has been rapidly increasing, and there is no sign of stopping. It leads to a severe shortage of test kits and accurate detection models. A recent study demonstrated that the chest X-ray radiography outperformed laboratory testing in COVID-19 detection. Therefore, using chest X-ray radiography analysis can help to screen suspected COVID-19 cases at an early stage. Moreover, the patient data is sensitive, and it must be protected to avoid revealing through model updates and reconstruction from the malicious attacker. In this paper, we present a privacy-preserving Federated Learning system for COVID-19 detection based on chest X-ray images. First, a Federated Learning system is constructed from chest X-ray images. The main idea is to build a decentralized model across multiple hospitals without sharing data among hospitals. Second, we first show that the accuracy of Federated Learning for COVID-19 identification reduces significantly for Non-IID data. We then propose a strategy to improve model's accuracy on Non-IID COVID-19 data by increasing the total number of clients, parallelism (client fraction), and computation per client. Finally, we apply a Differential Privacy Stochastic Gradient Descent (DP-SGD) to enhance the preserving of patient data privacy for our Federated Learning model. A strategy is also proposed to keep the robustness of Federated Learning to ensure the security and accuracy of the model.

%% file: introduction.tex

Coronavirus disease 2019 (COVID-19) caused by severe acute respiratory syndrome coronavirus 2 (SARS-CoV-2). It has spread worldwide and leading to the ongoing 2021 pandemic since the first known case has been identified in Wuhan, China, in December 2019. With more than 150 million confirmed cases and three million deaths across nearly 200 countries, COVID-19 is continuing to spread around the world, and there is no sign of stopping. Thus, it leads to a problematic situation for humans in the world until now. Although COVID-19 vaccines have provided an opportunity to slow the spread of the virus and end the pandemic, not enough COVID-19 vaccines are available for everyone in the world to be inoculated until the end of 2024 at the earliest, said by the chief executive of the world's largest vaccine manufacturer. Moreover, the appearance of mutations of new variants of the COVID-19 virus could make the virus more infectious or more capable of causing severe disease, for example, the latest variants of SARS-Cov-2 including at least two mutations in India.
The symptoms of COVID-19 often include fever, chills, dry cough, and systemic pain. However, there are large of mount people infected with the virus without noticeable symptoms. Thus, it leads to difficult-to-diagnose COVID-19 infections. Moreover, if the patient is detected early, it will be more likely to be cured faster and limit the disease's spread to others in the community. Therefore, it is urgent to find a method to help the hospital diagnose COVID-19 patients as soon as possible.


Many researchers have applied AI technology for building detection models to help hospitals detect patients as soon as possible. Most COVID-19 cases display common features on chest radiography images, including early ground-glass opacity and late-stage pulmonary consolidation. It can also be identified through a rounded morphology and a peripheral lung distribution~\cite{chung2020ct,huang2020clinical}. In 2020, research published in Journal Radiology ~\cite{wang2020temporal} demonstrated that chest radiography outperformed laboratory testing in detecting coronavirus. Therefore, using chest radiography image analysis can help to screen suspected COVID-19 cases at an early stage. In particular, chest X-ray image has various advantages like high accessibility, cheap, easy to operation, and it helps in the rapid prioritization of COVID-19 suspected patients.

Most of research use chest X-rays radiography (CXR), chest computed tomography (CT) and Lung ultrasound (LUS)~\cite{horry2020covid,gomes2020potential,khuzani2020covid} as screening methods. These methods heavily rely on shared datasets for the training process. However, based on General Data Protection~\cite{voigt2017eu}, patient data privacy must be protected to avoid an attack from the malicious attacker because data privacy direct impacts human politics, businesses, security, health, finances, etc. Therefore we need to find a better way so that machine learning can work collaboratively but still keep data privacy. One recent method that tackles this problem is Federated Learning (FL), proposed by Google~\cite{mcmahan2017communication}. Its main idea is to build a decentralized machine learning model based on datasets across multiple data sources without sharing data among the sources. The model updates focus more on the learning task than raw data, and the server only needs to hold individual updates ephemerally. With these features, FL offers significant privacy improvements compared to centralizing all training data. Several researchers have applied FL for COVID-19 detection tasks and achieved promising results~\cite{kumar2020blockchain,liu2020experiments,wang2020covid}. However, some works have demonstrated that FL may not always provide sufficient privacy guarantees. The sensitive information still can be revealed through model updates ~\cite{bhowmick2018protection, fredrikson2015model, melis2019exploiting}. For example, ~\cite{aono2017privacy} shows that the local data information can be revealed from a small portion of gradients, or a possible scenario is that the malicious attacker can reconstruct the training data from gradients information in a few an iterations~\cite{zhu2020deep}.

In summary, this paper makes the following contributions:
\begin{itemize}
\item We propose a privacy-preserving Federated Learning model for COVID-19 detection based on chest X-ray images collected from multiple sources (i.e., Hospitals) without sharing data among data owners by adding the Differential Privacy Stochastic Gradient Descent (DP-SGD)  resilient to adaptive attacks auxiliary information.
\item We demonstrate that the accuracy of Federated Learning for COVID-19 detection reduces significantly for Non-IID data due to varying size and distribution of local datasets among different clients. We thoroughly analyze several design choices (e.g., the total number of clients, amount of multi-client parallelism, computation per client) to improve the model's accuracy with Non-IID data.

\item We then provide a strategy to keep the robustness of our privacy-preserving Federated Learning model to ensure the model's security and accuracy.
\end{itemize}

%% file: background.tex
Since we aim to build a FL system to detect COVID-19 diseases with higher privacy by adding the Differential Privacy Stochastic Gradient Descent(DP-SGD), this section provides a comprehensive overview of FL and DP-SGD used in our COVID-19 system.
\subsection{Federated Learning}
The term Federated Learning (FL)  was introduced in 2016. It is a machine learning strategy where multiple clients can collaboratively solve a machine learning problem when each client stores its own data and will not exchange or transfer data with other clients~\cite{mcmahan2017communication}. The FL requires less storage or computational resources in the central server than centralize learning, and the most important is that FL helps protect each client's private data.

Since the term FL was initially implemented over a larger number of small devices ~\cite{brendan2918FL,mcmahan2017communication}. The various implemented FL applications have greatly increased, including some which might involve only small number of clients in collaboration among institutions~\cite{kairouz2019advances}. These two FL settings are called "cross-device" and "cross-silo" respectively. A typical FL training is achieved by following several basics steps. In the first step, all chosen clients download the current weight of the master model. Secondly, the clients compute the weight and update independently based on their local data. Finally, all clients update their weight to the server, where they are gathered and aggregated to produce a new master model. These steps are repeated until a certain convergence criterion is satisfied.

In our setting, we term the  FederatedAveraging (FedAvg)~\cite{mcmahan2017communication} as our FL system. In this way, for each communication round, the selected clients will compute the gradient of the loss on the current model using their local data. Then the server takes a weighted average of the resulting models. The pseudo-code of FedAvg is given in Algorithm 1.
\begin{algorithm}
	\label{alg:algorithmFedAvg}
	\caption{FederatedAveraging. The $K$ clients are indexed by $k$, $B$ is the local minibatch size, $E$ is the number of local epochs, and $\eta$ is the learning rate.}
	\textbf{Service executes:}
	\begin{algorithmic}
		\State $w_{0} \leftarrow$ random initialization
		\For { each round $t$ = 1,2,...}
		\State $S_{t} \leftarrow $ (random subset of $max(C\times K,1)$ clients)
		\For {each client $k \in S_{t}$ in parallel}
		\State $w_{t+1}^k \leftarrow$ ClientUpdate($k$,$w_t$)
		\EndFor
		\State $w_{t+1} \leftarrow \sum_{k=1}^K\frac{n_k}{n}w_{t+1}^k$
		\EndFor
	\end{algorithmic}
	\textbf{Client update($k,w$):}
	\begin{algorithmic}
        \State Split local dataset in $B$ ( $\frac{B}{n_k}$ batches of size $B$)
		\For{epoch $e \in\left[1,E\right]$}
		\For {batch $b \in B$}
		\State $w \leftarrow w - \eta\triangledown\ell(w;b) $
		\EndFor
		\EndFor
		\State return $w$ to server
	\end{algorithmic}
\end{algorithm}
\subsection{Differential Privacy Stochastic Gradient Descent (DP-SGD)}
Differential privacy is a strong standard for quantifying and limiting information disclosure about individuals~\cite{dwork2011firm,dwork2006calibrating,dwork2014algorithmic}. It aims to mask the contribution of any individual user by introducing a level of uncertainty into the released model. Differential privacy is quantified by privacy loss parameters ($\epsilon$, $\delta$). Where $\epsilon$ is how much a person with output would be able to see the dataset, $\delta$ is the probability that an unwanted event happens that leaks more data the normally. The smaller ($\epsilon$, $\delta$) corresponds to increased privacy. We have a differential privacy definition as following:
a randomized algorithm $A$: $D$ $\rightarrow$ $R$ with domain $D$ and range $R$ is ($\epsilon$, $\delta$)-differential private if for any subset of outputs S $\subseteq$ $R$ and for any two adjacent inputs $d, d'$ $\in$  $D$:
$Pr\left[A(d) \in S\right] \leq e^{\epsilon}Pr\left[ A(d') \in S\right] +\delta$.

In the term of Federated Learning, we say that two decentralized datasets $D$ and $D'$ are adjacent if they differ in a single entry, that is, if $D'$ can be obtained from $D$ by adding or subtracting all the records of a single client. $\delta$ is preferably smaller than $\frac{1}{\mid d\mid}$. 

\begin{figure}
\centering
\includegraphics[width=3.5in]{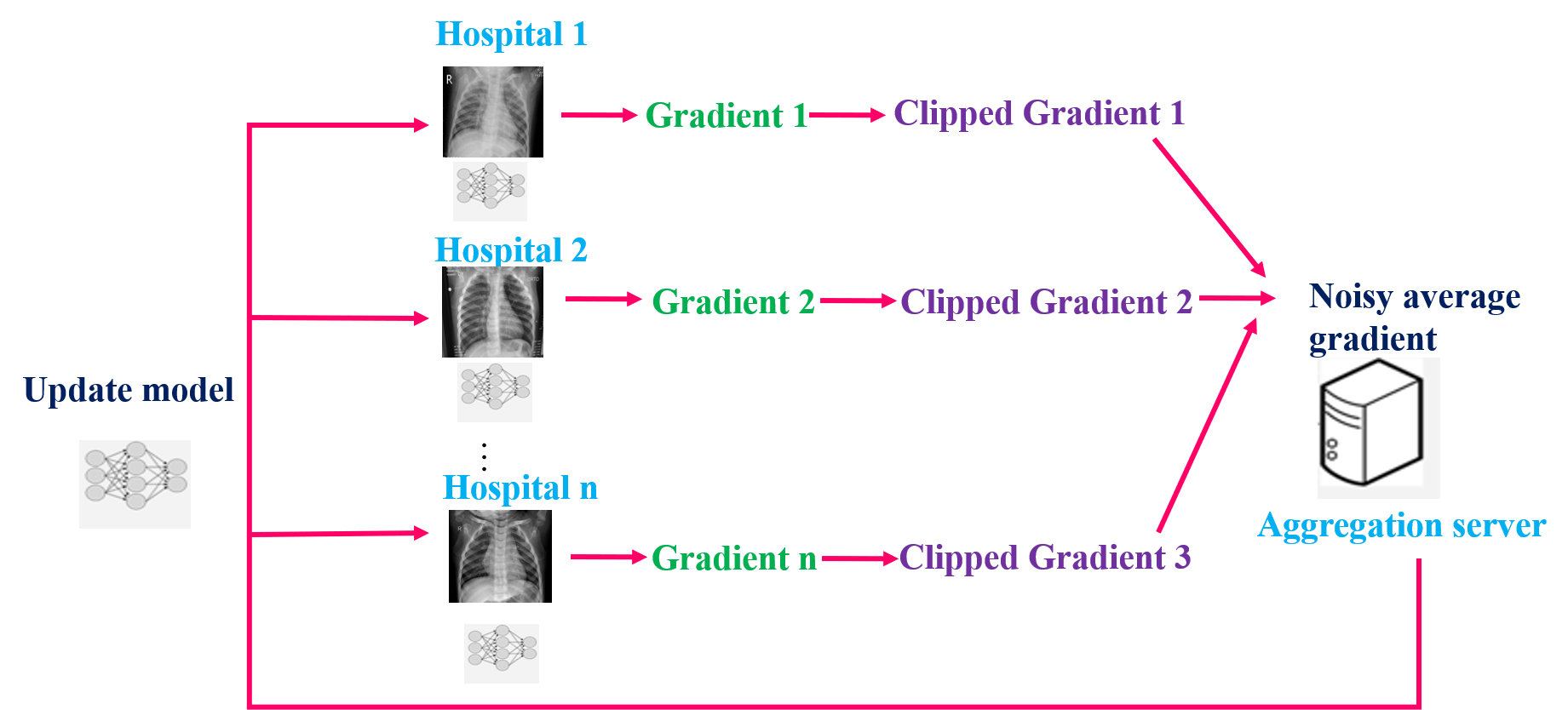}
\caption{The COVID-19 detection system architecture.}
\label{fig:dp_sgd}
\end{figure}

Differential privacy guaranteeing may impact the accuracy or utility of our model. In the rich data setting, it seems the model can enjoy both low privacy risk and high utility. However,  the optimization methods for large datasets must be scalable. Therefore, we utilize a Stochastic Gradient Descent to control the influence of training data during the training process as followed in previous works \cite{song2013stochastic, bassily2014private, abadi2016deep} for our differential privacy setting. The differential privacy stochastic gradient descent (DP-SGD) strategy is adding random Gaussian noise on the aggregated global model that is enough to hide any single client's update. It is consists of the following steps:  at each step of the DP-SGD, we compute the gradient for a random subset of examples, then we clip these per-sample gradients into a fixed maximum norm $\ell_{2}$. Next, random noise is added into the clipped gradients in computing the average step. Finally, we multiply these clipped and noised gradients by the learning rate and apply the product to update model parameters.
The setting of DP-SGD in our work is shown in Figure~\ref{fig:dp_sgd}.

%% file: experiments.tex
\subsection{Data collection and processing}
Following a team of researchers from Qatar University and the University of Dhaka, Bangladesh, along with their collaborators from Pakistan and Malaysia in collaboration with medical doctors~\cite{chowdhury2020can,rahman2021exploring}, we collected a dataset containing  3616 COVID-19 positives, 10192 normal and, 1345 viral pneumonia chest X-ray images. The COVID-19 data are collected from the various publicly accessible dataset, online sources, and published papers~\cite{mittal2021novel,Winther2020,Senzacategoria,eurorad,cohen2020covid,COVID19ChestXRayImageRepositoryfigshare,haghanifar2020covidcxnet}, normal data are collected from two different dataset~\cite{RSNAPneumoniaDetectionChallenge,ChestX-RayImages(Pneumonia)}, and viral pneumonia data are collected from chest X-ray images (pneumonia) database~\cite{ChestX-RayImages(Pneumonia)}. For each class, we randomly keep 200 images for testing data and the rest of the images for training.
\subsection{Federated Learning with different algorithm}
\label{subsec:Differentalgorithm}
\begin{figure}
\centering
\includegraphics[width=3.5in]{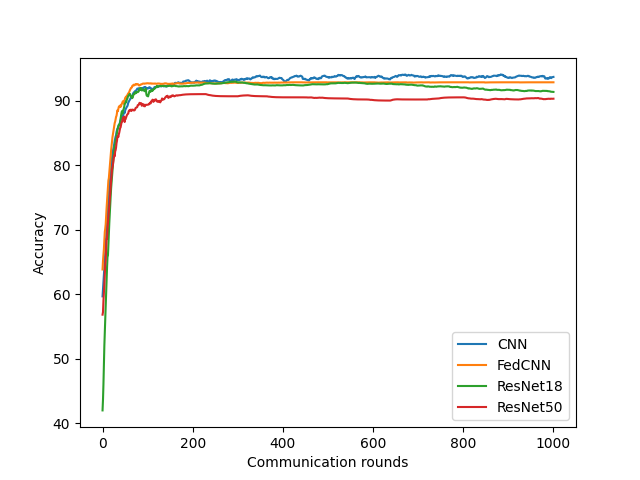}
\caption{Federated Learning with different alogrithm.}
\label{fig:FLDifferentAlogrithm}
\end{figure}

To choose a good model for our image classification task applying FL, we conduct experiments using four different models: (1) a convolutional neural network with two 3x3 convolutional layers, the first layer with 32 channels, and the second with 64 channels followed with 2x2 max pooling layer. A fully connected layer with 128 units and ReLu activation and a final softmax output layer. To reduce overfitting, we also add two dropout layers before and after the fully connected layer with the dropout probability of 0.25, 0.5 respectively. (2) a FedCNN model used in~\cite{mcmahan2017communication}, which includes two 5x5 convolution layers, the first convolution layer has 32 channels, the second layer has 64 channels, each layer followed with 2x2 max pooling, and the fully connected layer with 512 units and ReLu activations. (3) The popular classification model ResNeyt with 18 layers named ResNet18. (4) Resnet model with 50 layers named ResNet50.
We compare these four different models by running experiments with the number of client K = 3, client fraction C =0.33 corresponds to two clients per round, local epoch E = 1, client learning rate $\eta$ = 0.02, and the local minibatch size B = 20. As shown in Figure~\ref{fig:FLDifferentAlogrithm}, The CNN model achieves the best performance with 93.83\% accuracy after 1000 communication rounds. Therefore we will use this CNN model for our COVID-19 detection system.

\subsection{IID vs Non-IID}
\label{subsec:IIDvsNonIID}
Unlike centralized models, FL usually faces a Non-IID problem. The size and distribution of local datasets will typically vary heavily between different clients because each client corresponds to a particular user, a particular geographic location, or a particular time window.
In this work, we run experiments with both IID and Non-IID datasets and compare the performance of our model between these two kinds of datasets. For the IID setting, each client is randomly assigned a uniform distribution over all classes.
For the Non-IID setting, we first sort the data by the class label. We then divide data into two cases: (1) Non-IID(1), 1-class non-IID, where each client receives data partition from only a single class, and (2) Non-IID(2), 2-class Non-IID, where each client receives data partition from at most two classes.

\begin{figure}
\centering
\includegraphics[width=3.5in]{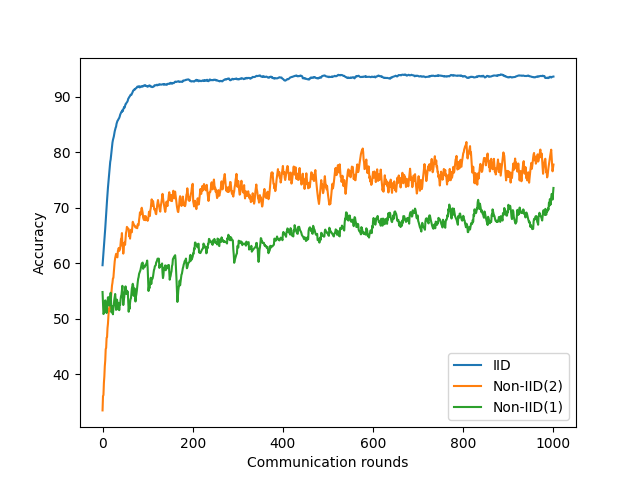}
\caption{IID vs Non-IID}
\label{fig:IIDvsNonIID}
\end{figure}

Similar to Section~\ref{subsec:Differentalgorithm}, the following parameters are used in our experiments: number of client K = 3, number of fraction C = 0.33 corresponds to two clients per round, local epoch E = 1, client learning rate $\eta$ = 0.02, and the local minibatch size B = 20. As shown in Figure~\ref{fig:IIDvsNonIID} and Table~\ref{table_IIDvsNonIID}, a significant reduction is observed on Non-IID data than IID data. The maximum accuracy reduction occurs for the most extreme 1-class non-IID(1), about 13 to 25\%. For 2-class non-IID(2), the accuracy reduction around 10 to 16\%. From these experiments, we see that non-IID data is one of the major issues of the FL system, and it is necessary to find a way that can help improve our model performance on non-IID data. In the next section, we will propose a strategy to improve our Non-IID(1) data performance.

\begin{table}[]
\centering
\caption{Performance comparison between IID vs Non-IID.}
\label{table_IIDvsNonIID}
\scalebox{1}{
\begin{tabular}{|l|l|l|l|}
\hline
Round & IID & Non-IID(1)  & Non-IID(2)
\\ \hline
400   & \textbf{92.67\%}    & 65.35\% & 83.50\%  \\ \hline
600  & \textbf{93.17\%}    & 65.83\% & 82.33\%  \\ \hline
800   & \textbf{93.69\%}    & 68.57\% & 79.87\% \\ \hline
1000   & \textbf{93.67\%}    & 73.59\% & 77.81\% \\ \hline
\end{tabular}}
\end{table}
\subsection{Non-IID improvement}
In this section, we evaluate different parameters in our Non-IID(1) setting to find the relationship between these parameters and our model performance. These parameters are the number of total clients K, client fraction C, and local mini batch-size B.
\subsubsection{Non-IID with different number of client}
\label{subsec:Noniidwithdifferentclient}
\begin{figure}
\centering
\includegraphics[width=3.5in]{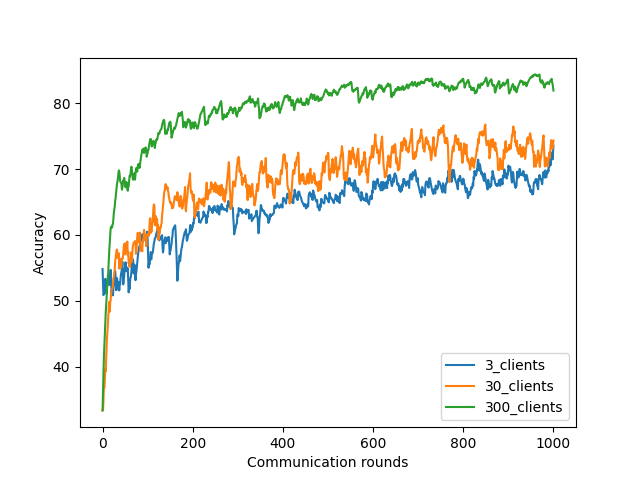}
\caption{Non-IID with different number of client.}
\label{fig:NonIIDdifferentclient}
\end{figure}
\begin{table}[]
\centering
\caption{Non-IID with different number of client.}
\label{table_NonIIDdifferentclient}
\scalebox{1}{
\begin{tabular}{|l|l|l|l|}
\hline
Round & 3 clients & 30 clients  & 300 clients
\\ \hline
400   &65.35\%    & 71.20\% &\textbf{80.23\%}  \\ \hline
600  & 67.73\%    & 71.80\% & \textbf{81.01\%}  \\ \hline
800   & 68.57\%    & 72.23\% & \textbf{83.72\%} \\ \hline
1000   & 73.59\%    & 74.36\% & \textbf{81.95\%} \\ \hline
\end{tabular}}
\end{table}
We first experiment on Non-IID with the various total number of clients (3, 30, 300) while keeping the other parameters:  client fraction C = 0.33, local epoch E = 1, client learning rate $\eta$ = 0.02, and the local minibatch size B = 20. Figure~\ref{fig:NonIIDdifferentclient} and Table~\ref{table_NonIIDdifferentclient} show the impact of varying K for our COVID-19 detection model. It demonstrated that using a larger number of client K = 30 and K = 300 show a significant improvement for our Non-IID setting compared to the model using K = 3. This can be explained by using a larger number of clients for our Non-IID(1) setting; some clients may receive data from the same class. Therefore our model can easily recognize the given clients' patterns if the model already learns similar data patterns in previous clients. This lead to improvement in model accuracy.
\subsubsection{Increasing client fraction}
\label{subsec:Noniidwithdifferentfraction}
\begin{figure}
\centering
\includegraphics[width=3.5in]{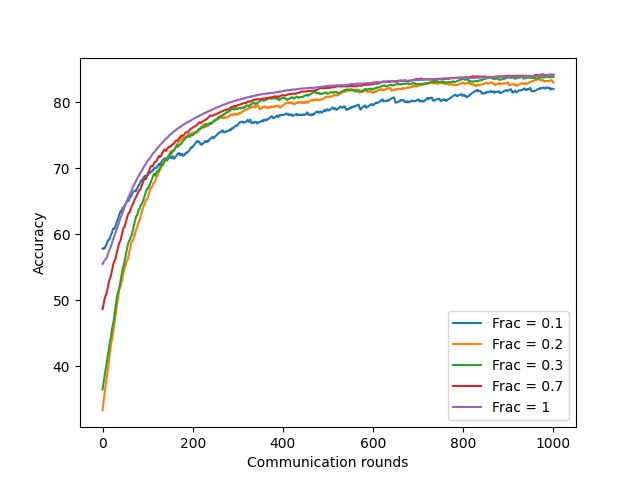}
\caption{Non-IID with different client fraction.}
\label{fig:NonIIDdifferentclientfraction}
\end{figure}

In this experiment, we evaluate our model with the client fraction C, which controls the amount of multi-client parallelism. To compute this, we fix local epoch E = 1, batch-size B = 20, client learning rate $\eta$ = 0.02, number of total client K = 300 (achieve best performance in previous Section~\ref{subsec:Noniidwithdifferentclient}), while changing the ratio of client fraction C with varying value $\in$ \{0.1, 0.2, 0.3, 0.7, 1\}. As shown in Figure~\ref{fig:NonIIDdifferentclientfraction}, using the larger client fraction shows improvement for our model accuracy. Moreover, with the same number of communication rounds, the model with a larger client fraction helps the model convergence faster.

\subsubsection{Increasing computation per client}
\label{subsec:Noniidwithdifferentbtachsize}
\begin{figure}
\centering
\includegraphics[width=3.5in]{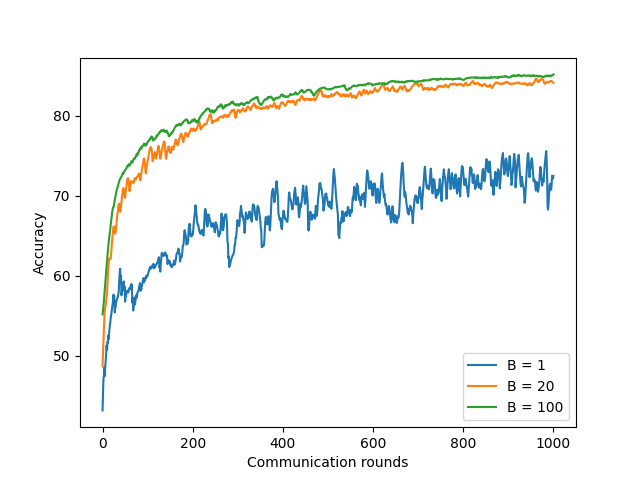}
\caption{Non-IID with different batch-size.}
\label{fig:NonIIDdifferentbatchsize}
\end{figure}

The last parameter we want to evaluate the affection on the Non-IID model is the local batch-size B. We fix client fraction C = 0.7, which shows the improvement results in the previous section, local epoch E = 1, client learning rate $\eta$ = 0.02, and the number of client K = 300, while the local batch-size value will be selected by varying value $\in$ \{1, 20, 100\}. Figure~\ref{fig:NonIIDdifferentbatchsize} shows that with 1000 communication rounds, the model with small batch-size B = 1 achieves the lowest accuracy around 72.47 \%, while larger batch-size B = 20 achieves improvement result with 84.11\% of accuracy, and the best result is achieved by the model using largest batch-size B = 100 with 85.18 \% of accuracy.
\subsubsection{NonIID vs refined NonIID}
\begin{table}[]
\centering
\caption{Non-IID with refined Non-IID.}
\label{table_NonIIDvsrefinednoniid}
\scalebox{1}{
\begin{tabular}{|l|l|l|l|}
\hline
Round & Base-line & Non-IID(300clients) & Refined Non-IID
\\ \hline
400  &65.35\% &80.23\%    &\textbf{82.59\%}  \\ \hline
600  &65.83\% &81.01\%    & \textbf{83.86\%}  \\ \hline
800  &68.57\% &83.72\%   & \textbf{84.50\%} \\ \hline
1000 &73.59\% &81.95\%  & \textbf{85.18\%} \\ \hline
\end{tabular}}
\end{table}
In previous Sections~\ref{subsec:Noniidwithdifferentclient}, ~\ref{subsec:Noniidwithdifferentfraction}, ~\ref{subsec:Noniidwithdifferentbtachsize}, we find out that each component has affection on model performance. We now combine these parameter values to see how much our model accuracy can be increased compared to base-line model. The base-line model is Non-IID(1) used in Section~\ref{subsec:IIDvsNonIID} with number of clients K = 3, fraction C = 0.33, client learning rate $\eta$ = 0.02, and local batch-size B = 20. Non-IID(300 clients) is a modified of the base-line model with number of clients K = 300. The final refined model is a refined model from Non-IID (300 clients) with number of client fraction = 0.7, and local batch-size = 100.
Table~\ref{table_NonIIDvsrefinednoniid} shows that using the number of clients K = 300 helps to improve accuracy by up to $\sim$ 15.18\% compared to the base-line model using three clients. Furthermore, we further improve the accuracy model of Non-IID(300 clients) by up to $\sim$ 3.23\% with an increasing client fraction and local batch-size.

\subsection{Federated Learning with DP-SGD}
Without sharing each client's private data, FL helps to mitigate privacy risks from centralized machine learning. However, the adversary might infer our information from the shared gradients from the previous model in our FL system. In order that an adversary is much more difficult to breach privacy, we apply Different Privacy Stochastic Gradient Descent (DP-SGD) by adding a random Gaussian noise on gradients on the aggregated model.

\begin{figure}
\centering
\includegraphics[width=3.5in]{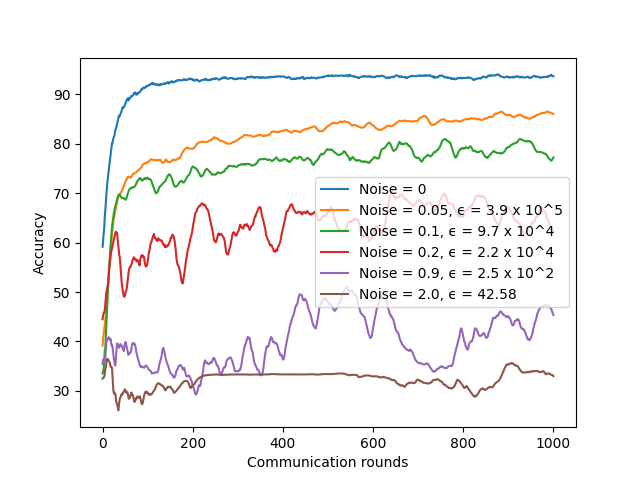}
\caption{Differential privacy with different noise value.}
\label{fig:DPdifferentnoisevalue}
\end{figure}

We first run experiments for IID dataset setting using number of total clients K = 3 with varying noise values to see how much differential privacy impacts to our utility model. We set $\delta$ = $10^{-5}$, client fraction = 0.33, client learning rate $\eta$ = 0.02, and batch-size = 20.
As shown in Figure~\ref{fig:DPdifferentnoisevalue}, we see a trade-off between our privacy and model accuracy. The more noise we add to our model, the more reduced our model accuracy is.

\begin{figure}
\centering
\includegraphics[width=3.5in]{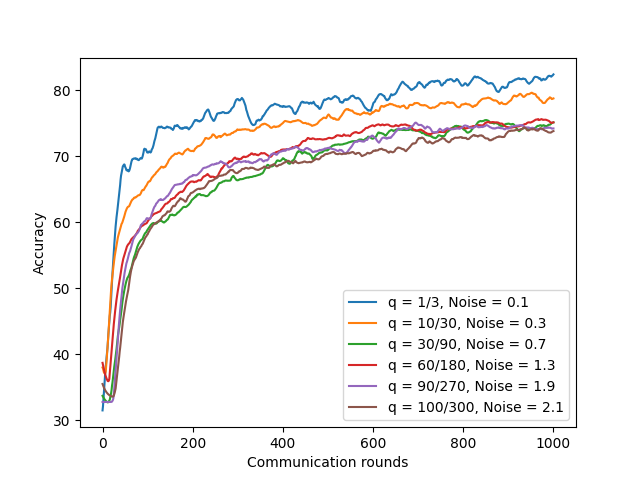}
\caption{The robustness of differential privacy.}
\label{fig:DPdifferentnumberclient}
\end{figure}
\begin{table}[]
\centering
\caption{The robustness of differential privacy.}
\label{table_DPdifferentnumberclient}
\scalebox{1}{
\begin{tabular}{|l|l|l|l|}
\hline
q & Noise & Accuracy& $\epsilon$
\\ \hline
1/3   &0.1  &\textbf{82.40}\%  &9.7 x $10^{4}$  \\ \hline
10/30  & 0.3  &78.76\%  &  8.9 x $10^{3}$ \\ \hline
30/90   &0.7  &75.13\% & 5.6 x $10^{2}$ \\ \hline
60/180  &1.3  &75.14\% & 97.39 \\ \hline
90/270   &1.9  &74.21\% & 46.36 \\ \hline
100/300  &2.1  &73.81\% & \textbf{39.40} \\ \hline
\end{tabular}}
\end{table}

Next, we try to find a way to lower the model privacy risk. Still, we somehow can keep the similar utility of our model by evaluating three system parameters: fraction of model q (number of client per round / total number of client), number of total clients, and the noise value. In this experiment, we will scale up the total number of clients while keeping the fraction of the model constant, and the noise value will be scaled up using varying values $\in$ \{0.1, 0.3, 0.7, 1.3, 1.9, 2.1\}. The lower $\epsilon$ value means higher security. As shown in Figure~\ref{fig:DPdifferentnumberclient} and Table~\ref{table_DPdifferentnumberclient}, with increasing the total number of the client while keeping client fraction constant, we add larger noise value, and the model most likely achieve a similar utility but higher privacy. For example, when increasing the number of clients from 180 to 270, we can increase the noise value from 1.3 to 1.9, thereby reducing a half $\epsilon$ value from 97.39 to 46.36 while the accuracy model only reduces 1\% from 75.14\% to 74.21\%. 

%% file: conclusion.tex
In this paper, we presented a privacy-preserving Federated Learning system for COVID-19 detection based on chest X-ray images. Through four different models: CNN, FedCNN, ResNet18, ResNet50, our Federated Learning applying CNN achieves the best performance with 93.83\%, and it is chosen for our COVID-19 identification experiments. We first show that the accuracy of Federated Learning for covid-19 identification reduces significantly by up to $\sim$ 25\% for Non-IID data. As a solution, we propose a strategy to improve accuracy on Non-IID data by increasing the total number of clients, parallelism (client fraction), and computation per client (batch size). Experiments show that model accuracy can be increased by $\sim$ 18.41\%. Second, to enhance patient data privacy for our Federated Learning model, we apply a Differential Privacy Stochastic Gradient Descent (DP-SGD) that is resilient to adaptive attacks using auxiliary information. Finally, we propose a strategy to keep the robustness of Federated Learning to ensure the security and accuracy of the model by keeping the sampling probability model, scaling up the total number of clients, and scaling up noise proportionally.

Although the proposed method is designed for COVID-19 detection, this method can detect other types of medical images as long as labeled images are available. In our future work, we would like to implement this method using a larger dataset available from various hospitals around the world. Furthermore, we hope that our proposed privacy-preserving Federated Learning framework enhances data protection for collaborative research to fight the COVID-19 pandemic. 